\documentclass[12pt,preprint]{aastex}
\usepackage{natbib,psfig}

\slugcomment{to appear in ApJ Letters}

\shorttitle{Symmetry in the changing jets of SS\,433}
\shortauthors{Blundell \& Bowler}

\begin{document}

\title{Symmetry in the changing jets of SS\,433\\ and its true distance from us} 

\author{Katherine M.\ Blundell\altaffilmark{1} and Michael G.\ Bowler
\altaffilmark{1}}

\altaffiltext{1}{University of Oxford, Department of Physics, Keble
  Road, Oxford, OX1 3RH, U.K.}

\begin{abstract}
We present the deepest yet radio image of the Galactic jet source,
SS\,433, which reveals over two full precession cycles ($> 2 \times
163$\,days) of the jet axis.  Systematic and identifiable deviations
from the traditional kinematical model for the jets are found:
variations in jet speed, lasting for as long as tens of days, are
necessary to match the detailed structure of each jet.  It is
remarkable that these variations are equal and opposite, matching the
two jets simultaneously.  This explains certain features of the
correlated redshift residuals found in fits to the kinematic model of
SS\,433 reported in the literature.  Asymmetries in the image caused
by light travel time enabled us to measure the jet speeds of
particular points to be within a range from $0.24\,c$ to $0.28\,c$,
consistent with, yet determined independently from, the speeds derived
from the famous moving optical emission lines.  Taken together with
the angular periodicity of the zigzag/corkscrew structure projected on
the plane of the sky (produced by the precession of the jet axis),
these measurements determine beyond all reasonable doubt the distance
to SS\,433 to be $5.5 \pm 0.2$\,kpc, significantly different from the
distance most recently inferred using neutral hydrogen measurements
together with the current rotation model for the Galaxy.

\end{abstract}


\section{Introduction}
\label{sec:intro}

SS\,433 is famous as the first known relativistic jet source in the
Galaxy.  Red- and blue-shifted optical lines, indicating velocities of
$0.26c$, were discovered by \citet{Mar79a,Mar79b} and interpreted as
being from gas accelerated in oppositely-directed jets
\citep{Fab79,Mil79}.  \citet{Mar84} successfully fitted a kinematic
precessing jet model, finding an intrinsic jet speed of
$\approx0.26c$, a precession period of $\sim 163\rm\,days$, a cone
opening angle of $\sim20^\circ$, and an inclination to the
line-of-sight of the precession axis of $\sim80^\circ$.  Subsequent
radio imaging was consistent with this model
\citep[e.g.,][]{Hje81,Ver87,Ver93}, with the images of \cite{Hje81}
resolving ambiguities in the parameters from optical studies.

While optical spectroscopic data observed over many precession cycles
\citep[e.g.][]{Mar84,Eik01} have confirmed the basic parameter values
of what has come to be known as the kinematic model, a detailed
analysis of any deviations of the data from this model is hindered by
inherent degeneracies in the parameters to be fitted, since the
measurement is that of line-of-sight Doppler shifts.  Detailed
analyses from quasi-daily milli-arcsec scale monitoring are similarly
afflicted by such degeneracies since, for example, a variation in
observed proper motion can in principle arise from changes in jet
velocity projected on the plane of the sky, or changes in the angle
the jet axis makes with our line-of-sight, perhaps caused by a change
in precession rate.  Moreover, in milli-arcsec scale monitoring only a
limited fraction of the precession period of 162 days has been sampled
to date.  However, the extended and spatially resolved jet output
observed over several arcseconds, while not time-resolved in the
conventional sense, is an historical record of the geometry of the jet
ejection over two complete precession periods.

\section{The deepest yet image of SS\,433}
\label{sec:obs}
On 2003 July 10, SS\,433 was observed for 10\,hrs with the VLA in its
most extended A-configuration.  The primary flux calibrator was
3C\,286 and the secondary calibrator was 1922+155.  The data were
reduced using standard techniques within the AIPS software and the rms
background noise was 29\,$\mu$Jy/beam.  The resulting total intensity
image is shown in Fig\,\ref{fig:stdmodel}a and reveals two oppositely
directed jets, each of which can be followed over more than 2 complete
precession periods.  The compression to a zigzag on the sky of the
eastern (mainly approaching) jet and corkscrewing of the western
(mainly receding) jet, both due to the finite speed of light, are
remarkably clear.  Superimposed on the image is the path traced on the
sky in the simple kinematic model of \cite{Hje81}, assuming a jet
speed of $0.26\,c$, period 162 days and a distance of 5.5\,kpc.  The
match in Fig\,\ref{fig:stdmodel}a is sufficiently good for it to be
obvious that if the mean speed is $0.26\,c$ and the period 162 days
then the distance to SS\,433 is 5.5\,kpc with an uncertainty of
perhaps $\pm 0.1\,$kpc.  If the true distance were 3.1\,kpc
\citep{Dub98} the path traced on the sky would, for a speed of
$0.26\,c$, have almost twice the angular periodicity.  The image could
only be (approximately) matched for a mean speed of $\sim
0.15\,c$. Then the bolides would have had to decelerate from $0.26\,c$
\citep{Eik01} and in doing so have lost $\sim 20\,$MeV per hydrogen
atom, two thirds of the original bulk energy, in a distance less than
$\sim 10$ light days. We regard this as so implausible that one might
conclude without further analysis that the distance to SS\,433 is
$\sim 5.5\,$kpc.  However, we establish this distance below {\em
without} assuming the mean speed.

\section{The true distance to SS\,433}
\label{sec:distance}
\begin{figure}
\epsscale{0.7}
\plotone{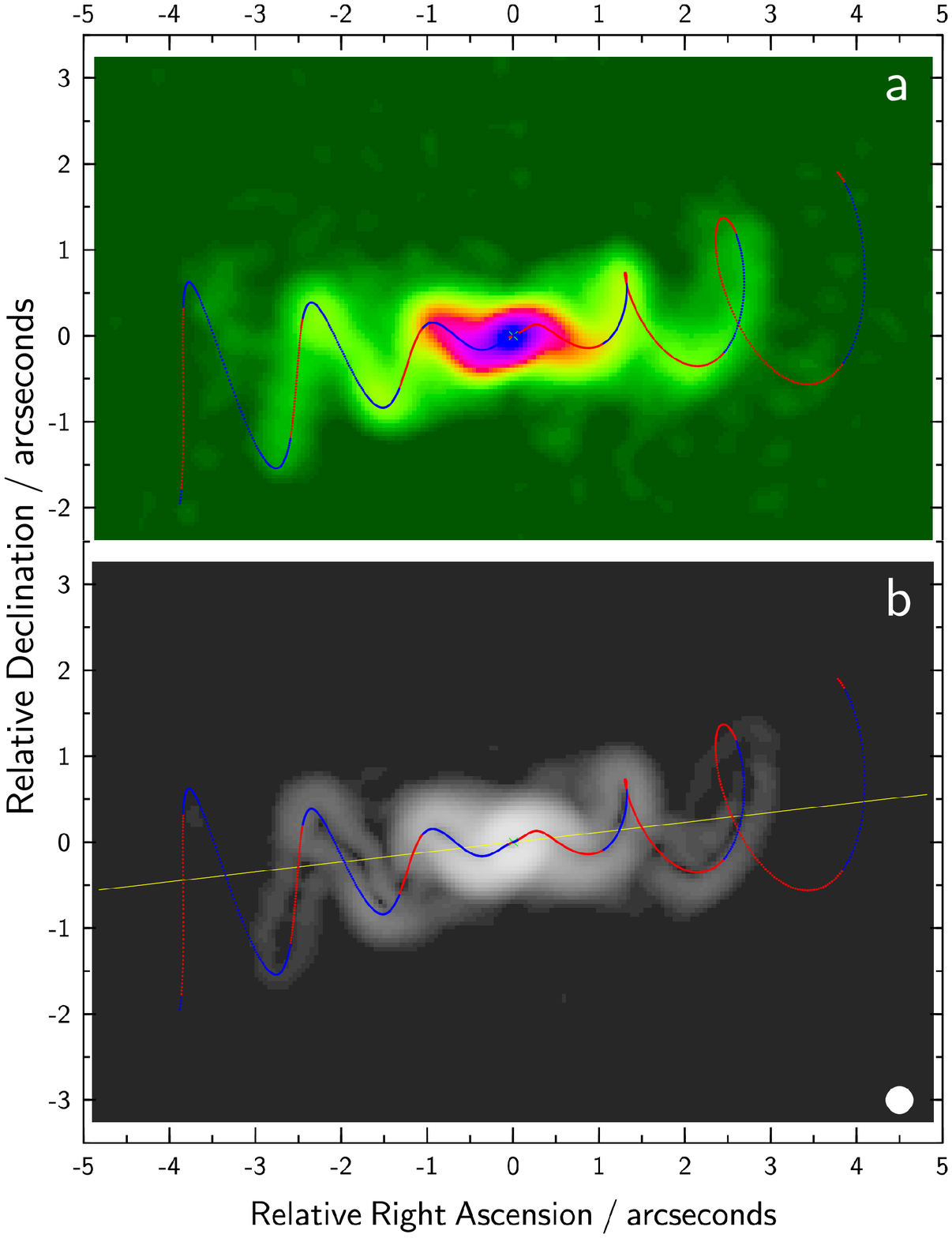}
\figcaption{\label{fig:stdmodel} {\bf (a)} A total intensity image at
  4.85\,GHz observed with the VLA in A-configuration.  Overlaid is the
  projection on the sky of two oppositely directed jets (composed of
  individual bolides ejected one per day in each direction) a distance
  5.5\,kpc from us, having constant ejection speed $0.26\,c$, whose
  axis precesses every 162 days and is at an angle of 78$^{\circ}$ to
  our line-of-sight (making an angle of 10$^{\circ}$ with respect to
  East-West) and tracing out a cone of semi-angle 19$^{\circ}$.  The
  blue regions of the line, indicate the bolides (whose motion is
  assumed to be ballistic) which are travelling towards us and the red
  regions indicate the bolides moving away from us.  To first order,
  the kinematic model seemingly shows a reasonable average fit to the
  data. {\bf (b)} A Sobel-filtered version of the image shown in (a)
  reveals more clearly the inadequacies of the standard kinematic
  model with respect to the data.  Implementation of the nodding
  parameters quoted by \cite{Sti02} makes only a modest difference to
  the appearance of the trace from the simple kinematic model.  The
  white circle in the lower right corner is the size of the
  $0.35^{\prime\prime} \times$ $0.35^{\prime\prime}$ synthesized beam.
}
\end{figure}

Close inspection of Fig\,\ref{fig:stdmodel}a shows that the real trace
departs from the simple kinematic model significantly.  This is
clearer in Fig\,\ref{fig:stdmodel}b.  The image shown is that of
Fig\,\ref{fig:stdmodel}a but with a Sobel operator having been
applied to delineate the ridgeline.  The Sobel operator \citep{Sob90}
is a simple edge detect filter which is widely used in magnetic
resonance imaging of brains.  The algorithm generating the filter
computes the square root of the sum of the squares of the directional
derivatives in two orthogonal directions.  The ridgeline shows up as
dark since the gradient changes from positive to negative across the
jet profile.

Distortions in the image introduced by the finite speed of light are
independent of distance so in the absence of such detailed departures
as are evidence in Fig\,\ref{fig:stdmodel}, the jet speed and hence
the distance would be extremely accurately determined.  Nonetheless,
the speed can be determined quite accurately under the assumptions
that the jets are symmetric and that special relativity is correct:
for a pair of bolides launched simultaneously but ejected in opposite
directions, each having velocity of magnitude $\beta$ in units of $c$
and making an angle $\theta$ to our line-of-sight, the ratio of their
separations from the central core is given by
\begin{equation}
  \frac{S_{\rm app}}{S_{\rm rec}} = \frac{1 + \beta \cos\theta}{1 -
  \beta \cos\theta},
\end{equation}
where $S_{\rm app}$ is the separation of the bolide which is
approaching us from the central core projected on the plane of the sky
and $S_{\rm rec}$ is the same for the receding bolide.  The validity
of the assumption of symmetry is established in
\S\,\ref{sec:deviations}.

A comparison of the separations from the central core to opposite
pairs of well-defined points where the ridgeline crosses the mean jet
axis (shown as the yellow line in Fig\,\ref{fig:stdmodel}b) yielded
speeds of $0.24 \pm 0.015\,c$, $0.25 \pm 0.03\,c$, $0.27 \pm 0.04\,c$
and hence a distance of $5.5 \pm 0.2$\,kpc.  Note that this is {\em
independent} of assumptions from optical data, in contrast with
previous estimates.  This value of the distance is rather larger than
that derived from quasi-daily milli-arcsec monitoring \citep{Sti02},
which is vulnerable to temporary deviations from the kinematic model,
which we explore in \S\ref{sec:deviations}, during observation,
together with the degeneracies mentioned in \S\ref{sec:intro}.

A distance of $5.5 \pm 0.2$\,kpc is somewhat above the weak upper
limit of 3.8\,kpc for the distance to SS\,433 which \cite{van80} found
from HI absorption measurements (together with a strong lower limit of
3.0\,kpc).  $5.5 \pm 0.2$\,kpc is substantially greater than the
distance of 3.1\,kpc inferred from neutral hydrogen measurements by
\cite{Dub98}.  They suggested that a gas cloud, seen in HI emission,
is interacting with W50: the inferred velocity of this cloud implies a
distance of 3.1\,kpc if the rotation model of the Galaxy is correct
(SS\,433's Galactic co-ordinates are [39.69, $-2.24$]).  However, this
model assumes purely circular motion and takes no account, for
example, of the presence of a bar in the Galaxy on the motion of gas
near SS\,433 \citep{Bin91}.  A distance of 5.5\,kpc for SS\,433
implies the same distance to the W50 nebula.

\section{Deviations from the kinematic model}
\label{sec:deviations}

\begin{figure}
\epsscale{0.8}
\plotone{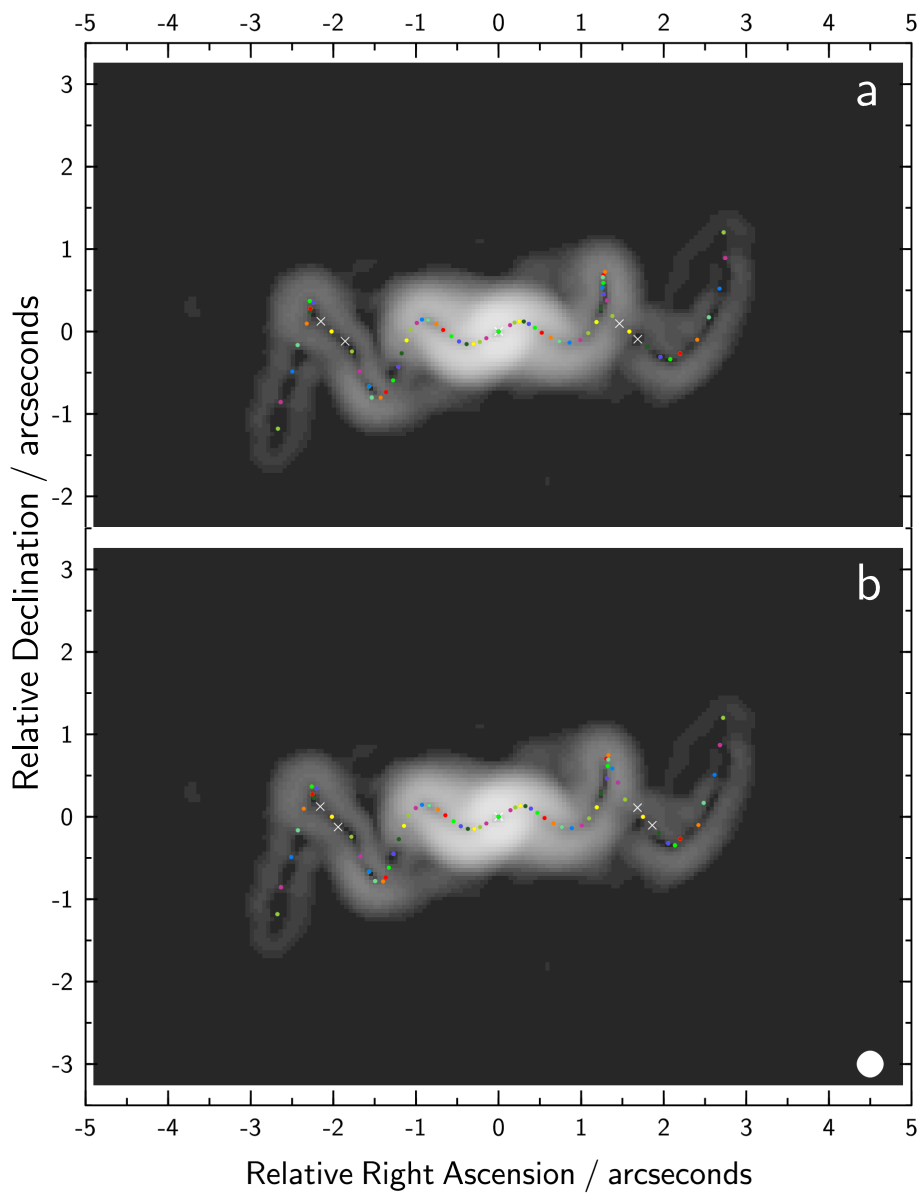}
\figcaption{\label{fig:sobel5p5} As Fig~\ref{fig:stdmodel}b, with the
image having been passed through a Sobel filter.  The coloured beads
(now representing one bolide emitted every 10 days in each direction
along the ambient jet axis) are the best by eye fits if jet speed is
the only varying parameter in the kinematic model and the distance of
SS\,433 from us is {\bf (a)} 5.5\,kpc requiring speeds which vary
around $0.26\,c$ and {\bf (b)} 3.1\,kpc requiring speeds which vary
around $0.15\,c$.  White crosses represent the positions of bolides
launched 235 and 245 days previously.  Details in main text.  }
\end{figure}

To verify the assumption of symmetry and investigate deviations from
the simple kinematic model, the distance was set to 5.5\,kpc and
simulated bolides were launched every ten days with equal speeds in
each jet, but the speeds were chosen to match the ridgeline of the
east jet.  This is illustrated in Fig\,\ref{fig:sobel5p5}a, where the
beads are colour coded such that matching colours on opposite sides
correspond to bolides launched simultaneously.  The beads shown for
the west jet are the symmetric counterparts of the eastern beads.  The
assumption of symmetry is very well justified over the entire image,
and outside the innermost precession period the test has precision.
(Note that although the FWHM of the point spread function of these
images is $0.35^{\prime\prime}$, the accuracy with which the centroid
of the peak of the jet cross-section may be found (and hence the jet
ridgeline known) is $\sim 0.035^{\prime\prime}$.)  The astonishing
accuracy with which the western jet is reproduced for a distance of
5.5\,kpc validates the assumption of symmetry, our delineation of the
ridgeline and our distance to SS\,433.  Fig\,\ref{fig:sobel5p5}b shows
the best fit that can be found for the eastern jet if the distance to
SS\,433 is 3.1\,kpc.  The beads shown for the west jet are the mirrors
of those shown for the east jet; inspection of
Fig\,\ref{fig:sobel5p5}b reveals that this does a very poor job of
fitting the west jet (quantified in Table\,\ref{tab:chisq}) and is
wholly inconsistent with jet symmetry.  As a consistency check on the
measurements of Table\,\ref{tab:chisq}, we calculated $\chi^2$ for the
east-jet beads at an assumed distance of 3.1\,kpc against the east-jet
beads at an assumed distance of 5.5\,kpc (this was 19.9) and for the
west-jet beads at an assumed distance of 3.1\,kpc against the west-jet
beads at an assumed distance of 5.5\,kpc (this was 110.3).

\begin{deluxetable}{lrr}
\tablecaption{\label{tab:chisq} To quantify the discrepancy between the fit
of the west jet to the ridgeline taking the fitted east jet values and
assuming symmetry, an independent smooth line was drawn along the
ridgeline of the jet, which was determined in most places to
$0.035^{\prime\prime}$.  With this uncertainty we then calculated the
$\chi^2$ as tabulated below for the sensitive outer regions. }
\tablehead{\colhead{Measure  } & \multicolumn{2}{c}{Distance}  \\
                   & \colhead{5.5\,kpc (Fig.\,\ref{fig:sobel5p5}a)}
                   & \colhead{3.1\,kpc (Fig.\,\ref{fig:sobel5p5}b)}\\ }
\startdata
$\chi^2$ for East jet (20 beads) & 16.2\phantom{xxxxxxxx} &  17.3\phantom{xxxxxxxx}  \\
$\chi^2$ for West jet (22 beads) & 17.3\phantom{xxxxxxxx} &  140.1\phantom{xxxxxxxx}  \\
\enddata
\end{deluxetable}

\begin{figure}
\epsscale{0.9}
\plotone{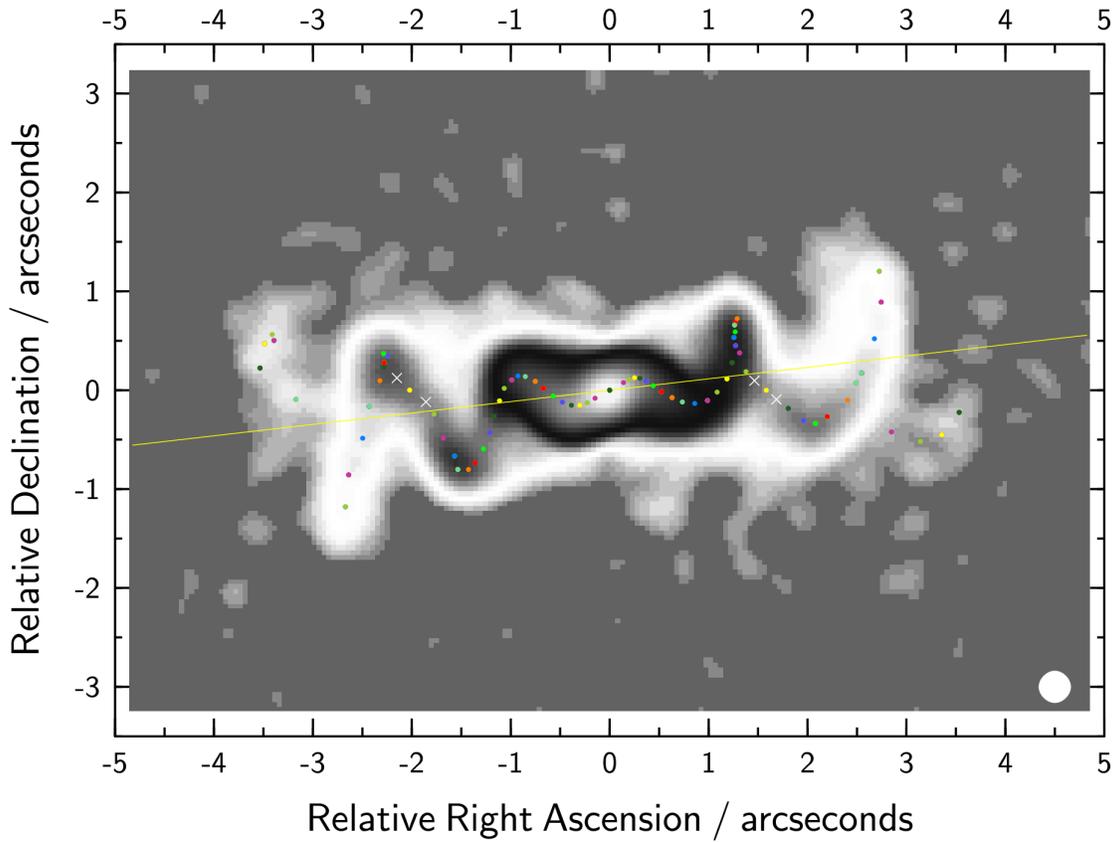}
\figcaption{\label{fig:finalfit} The same total intensity image as
Fig\,\ref{fig:stdmodel}a but shown with a transfer function which is
not monotonic, in order to emphasise local gradients in intensity over
a range of intensity levels and hence the good fit of the beads, which
are those of Fig\,\ref{fig:sobel5p5}a with the addition of a few
outside $\pm 3^{\prime\prime}$ relative Right Ascension. }
\end{figure}

\begin{figure}
\epsscale{0.9}
\plotone{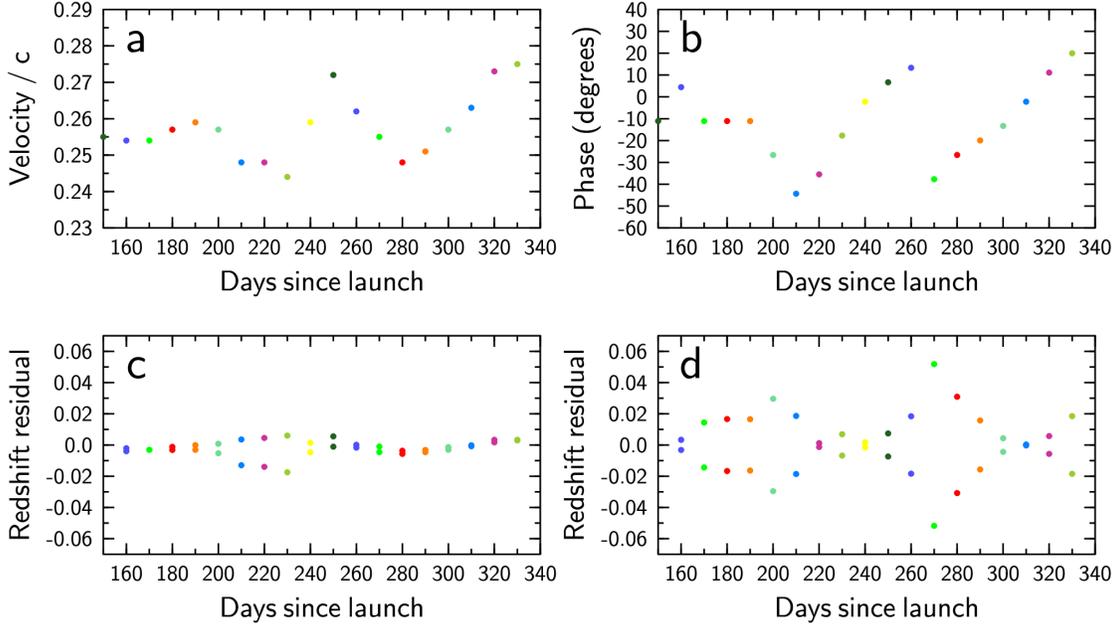}
\figcaption{\label{fig:redshifts} {\bf (a)} The velocities of
individual bolides versus the number of days since those bolides would
have been observed to be at the core, if our structural deviations are
purely attributable to variations in jet speed.  The colour coding is
the same as in Fig\,\ref{fig:sobel5p5}a.  {\bf (b)} The phases of the
jet axis as individual bolides were launched versus the number of days
since those bolides would have been observed to be at the core, if our
structural deviations are purely attributable to phase variations in
the precession of the jet axis.  Panels (c) and (d) pertain to the
lower panels of figures 2 and 3 of \cite{Eik01}: {\bf (c)} The
difference in redshifts which would have been observed had the
kinematic model been obeyed exactly and the redshifts which would have
been observed as the bolides were launched as in the scenario of (a).
{\bf (d)} As (c), but for the scenario of (b).  }
\end{figure}

Fig\,\ref{fig:finalfit} is a different rendering of the total
intensity image shown in Fig\,\ref{fig:stdmodel}a, with the beads
from Fig\,\ref{fig:sobel5p5}a (5.5\,kpc distance) superimposed, to
clearly demonstrate the consistency between the total intensity image
and that which had the Sobel operator applied.  The jet speeds vary
between $0.243\,c$ and $0.275\,c$.  These are shown in
Fig\,\ref{fig:redshifts}a, with the corresponding Doppler redshifts
for (hypothetical) optical spectroscopy shown in
Fig\,\ref{fig:redshifts}c plotted as residuals with respect to the
kinematic model.  These residuals are very modest compared with those
presented in \cite{Eik01}.

There is nothing in the accumulated data on optical Doppler shifts
\citep[e.g.,][]{Mar84,Eik01} which contradicts our attribution of the
observed departures from the simple kinematic model to these small
variations in velocity. On the contrary, there is in fact independent
evidence of speed variations in the spectroscopic data presented by
\cite{Eik01}. The principal features of the correlation between their
$z1$ and $z2$ are readily explained by (symmetric) velocity variations
in addition to symmetric pointing angle variations (as is the lack of
marked correlation with phase); we will discuss this in more detail in
a forthcoming paper (Blundell \& Bowler, submitted, see astro-ph).

The same symmetric match to the radio image may be obtained by keeping
the jet speed constant at $0.26\,c$ and varying the phase of the
precession.  The image on the sky is less sensitive to the phase than
is optical spectroscopy; the necessary phases are shown as a function
of time in Fig\,\ref{fig:redshifts}b, with the corresponding Doppler
residuals shown in Fig\,\ref{fig:redshifts}d.  The phase (in the
outer regions) varies approximately linearly with time, corresponding
to a rotation period of 276 days, rather different from the mean
period of 162 days.  There is also a phase jump of $\sim 60^{\circ}$
in less than 10 days.  The corresponding Doppler residuals are much
larger than those reported by \cite{Eik01}, precluding phase-only
variations being responsible for the observed deviations from the
kinematic model (unless our radio image captured a highly anomalous
episode).

We suggest that the  systematic deviations from the kinematic
model observed in optical spectroscopy arise from a combination of
both strictly symmetric oscillations in jet speed and strictly
symmetric variations in pointing angle.  

\section{Concluding remarks}
\label{sec:conclusions}

We have presented the deepest yet radio image of SS\,433 on arcsecond
scales, from which we have derived its distance from us to be
5.5\,kpc, independently of any assumptions from optical data.  We have
identified deviations from the simple kinematic model in our image,
which last over timescales of 10s of days.  These may be formally
fitted in either of two extreme scenarios: by small oscillations
purely in jet velocity about a mean of $0.26\,c$ (which are perfectly
symmetric in both jets) or by large oscillations purely in the rate of
precession about a mean of 162 days (which are perfectly symmetric in
both jets).  Additional information from redshift residual data
\citep{Eik01} strongly suggests that variations in velocity and in
phase are both occurring.

We speculate that the phase (precession rate) variations might arise
because of a varying effective moment of inertia of the nozzle (which
in turn might arise because of variations in the mass transfer rate).
We suggest that these same changes in mass distribution might cause
variation in the inner radius of the accretion disc which may determine
the speed of jet bolide ejection \citep{Mei01}.

\acknowledgments

K.M.B.\ thanks the Royal Society for a University Research Fellowship.
We warmly thank Rob Fender and Simone Migliari for suggesting deep
radio imaging of this source as part of a collaborative programme to
investigate the related X-ray and radio emission.  The VLA is a
facility of the NRAO operated by AUI, under co-operative agreement
with the NSF.  It is a pleasure to thank James Binney and Philipp
Podsiadlowski for useful discussions.

\newpage

\newpage

\newpage
\newpage

\end{document}